\documentclass[twocolumn,showpacs,amsmath,amssymb]{revtex4-1}

\usepackage{graphicx}
\usepackage{dcolumn}
\usepackage{bm}
\usepackage{epsfig,amsmath}

\def\vec#1{{\bf #1}}

\begin{document}

\title{Stability of low-friction surface sliding of nanocrystals with rectangular symmetry and application to W on NaF(001)}

\author{Astrid S. de Wijn}
\email{A.S.deWijn@science.ru.nl}
\affiliation{Radboud University Nijmegen, Institute for Molecules and Materials, Heyendaalseweg 135, 6525 AJ Nijmegen, The Netherlands}
\author{Annalisa Fasolino}
\affiliation{Radboud University Nijmegen, Institute for Molecules and Materials, Heyendaalseweg 135, 6525 AJ Nijmegen, The Netherlands}

\begin{abstract}
We investigate the stability of low-friction sliding of nanocrystal with rectangular atomic arrangement on rectangular lattices, for which analytical results can be obtained.
We find that several incommensurate periodic orbits exist and are stable against thermal fluctuations and other perturbations.
As incommensurate orientations lead to low corrugation, and therefore low friction,
such incommensurate periodic orbits are interesting for the study of nanotribology.
The analytical results compare very well with simulations of W nanocrystals on NaF(001).
The geometry and high typical corrugation of substrates with square lattices increase the robustness compared to typical hexagonal lattices, such as graphite.
\end{abstract}


\maketitle

\section{Introduction}

Friction at the atomic scale is currently actively studied \cite{Shirmeisen2009}.
One of the goals of this research is to understand whether extremely 
low friction can be obtained by an appropriate choice of the sliding conditions. 
Commensurability between the sliding lattices is one of the elements that determine friction.
For a purely incommensurate infinite contact, theoretical arguments suggest that static friction should vanish \cite{Aubry}.
However, very low friction has been measured also for finite incommensurate contacts at very low velocities \cite{Dienwiebel2004,Dienwiebel2008} and this effect has been called superlubricity \cite{Shinjo,flake,Consoli}.
The atomic force microscope (AFM) study \cite{Dienwiebel2004,Dienwiebel2008} found that the sliding of graphite flakes on graphite can occur with very low friction, depending strongly on the relative orientation.
At the same time, rotation of the flake can lead to a rapid increase of friction and stick-slip motion, corresponding to a locking into a commensurate orientation.
States of very different friction have also been observed to coexist in the sliding of nanoparticles \cite{Dietzel2008} and have been attributed to contamination or amorphous surfaces.

This paper examines theoretically the sliding of nanocrystals and substrates which both have different, but regular rectangular lattices.
By means of a simple, analytically soluble model, we show that stable orientations exist for any size nanocrystal, and derive some general properties, showing how the stability depends on the scan line.
The stable orientations are independent of the velocity and nearly independent of the corrugation, but depend only on the geometry of the substrate and contact layer of the nanocrystal.
We can also estimate the energy barriers necessary to rotate the nanocrystal from a given orientation to another.

We apply the results to W nanocrystals on an NaF(001) substrate.
NaF has been studied extensively as a substrate in the context of nanotribology
\cite{Hoelscher1996,Morita}
and W is commonly used for AFM tips (see, for instance Ref.~\cite{Dienwiebel2004}).
Both materials have a bcc lattice structure and the (001) surfaces in contact can thus be described by one finite and one infinite square lattice.
The lattice parameters of the two materials are very similar (0.31585~nm and 0.32668~nm respectively), which ensures the existence of nearly commensurate orientations for small enough nanocrystal.
For W, the corrugation on NaF(001) is of the order of 1~eV, a value typical for many substrate systems (see, for instance Ref.~\cite{Hoelscher1996}).

In this paper, we apply a model previously introduced to study the dynamics of hexagonal flakes on hexagonal lattices\cite{flake} to rectangular contact layers on substrates with rectangular lattices.
One of the important results is that the inherent lack of robustness of the low friction motion found in the hexagonal graphite system for particular scan lines is not present for rectangular lattices, which opens the possibility for practical applications of low-friction sliding.
The analytical results of the simplified model are supported by our numerical simulations of W nanocrystal on NaF.

In Sec.~\ref{sec:simple} we introduce the notation while briefly reviewing the model used in Ref.~\cite{flake} to describe the relation between the rotational dynamics and friction.
In Sec.~\ref{sec:UW}, this model is applied to a general system consisting of a rectangular lattice and a nanocrystal with a rectangular contact layer.
The analytical results are compared to numerical simulations of W nanocrystal on NaF(001) in Sec.~\ref{sec:simulations} and finally the implications of our results are discussed in Sec.~\ref{sec:discussion}.

\section{The simplified model\label{sec:simple} of friction and rotation}

\begin{figure}
\epsfig{figure=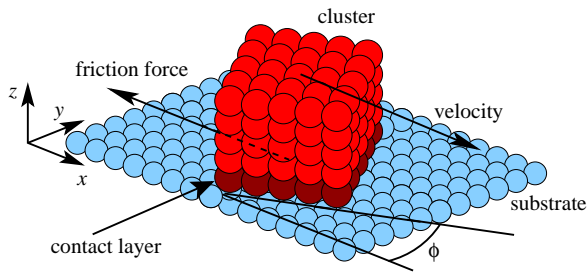,width=7.8cm}
\caption{
A schematic representation of a nanocrystal on a substrate.
Only the bottom layer (dark) of the nanocrystal interacts with the substrate.
\label{fig:nanocrystal}
}
\end{figure}

We wish to study the dynamics of small rigid nanocrystals of atoms arranged in a regular lattice on a regular substrate, as sketched in Fig.~\ref{fig:nanocrystal}.
The contact layer consists of one atomic plane and lies in the $x-y$ plane parallel to the periodic substrate.
For matched lattice parameters, by changing the orientation $\phi$ of the nanocrystal on the substrate, the contact layer can be either commensurate or incommensurate with the substrate atoms.
The incommensurate states which produce low friction can be destroyed by rotations around the $z$ axis that lead to a locking in a commensurate orientation \cite{Dienwiebel2008}.

Each atom of the contact layer is subjected to a periodic potential due to the substrate and to an external load force which may be applied to the nanocrystal.
Additionally, the contact layer atoms experience a friction force due to dissipation of kinetic energy into phonon modes of the substrate crystal. 
The remaining nanocrystal atoms are further away from the substrate and we may therefore assume that they couple only to the other nanocrystal atoms.
The AFM cantilever or a similar device used to investigate friction can be modelled by coupling the centre of mass of the nanocrystal harmonically with a spring with force constant $c$ to a support moving at constant velocity $\vec{v}_\mathrm{s}$.
The friction is given by the average value of the lateral force $F_\mathrm{s}$, exerted by the spring on the nanocrystal.

In three dimensions, a rigid nanocrystal is thus left with only 6 degrees of freedom: the coordinates of the centre of mass, and the orientation.
Despite this, the system is still too complicated to perform the stability analysis analytically.
In Ref.~\cite{flake} a simple model was proposed to describe the rotation of hexagonal graphite flakes on hexagonal graphite lattices and its consequences for low-friction sliding.
We briefly summarise this model and some of the results of Ref.~\cite{flake} here in a general context.

A possible simplification of the system is suggested by the nature of the dynamics.
We are interested in the rotation around the $z$ axis, which affects the commensurability, and therefore the friction.
Hence, the two relevant degrees of freedom are the position of the centre of mass along the scan line, $x$, and the orientation, $\phi$.
Consequently, instead of the full substrate potential $V(x,y,z)$, we introduce a potential $V(x,\phi)$, one example of which is given in Fig.~\ref{fig:V(x,phi)}, where one can see that the corrugation felt by the nanocrystal as a whole decays quickly away from the commensurate orientations $\phi=0,90^\circ$. 
Such a model is fully described by the initial support position $x_{\mathrm{s}}^0$, support velocity $v_\mathrm{s}$,
mass $M$ and moment of inertia $I$ of the nanocrystal, the effective potential $V(x,\phi)$, and the viscous friction coefficient $\gamma$ of a contact layer atom on the substrate.
Altogether, this model is similar to the Tomlinson model\cite{Tomlinson1929}, but with one additional degree of freedom which accounts for the rotational dynamics.
The details of the substrate and contact layer lattice geometry, choice of scan line, and the applied external load force are accounted for in $V(x,\phi)$ \cite{flake}.

We write the equations of motion of the simplified system as a dynamical system of first-order differential equations for the position $x$, velocity $v_x$, orientation $\phi$, and angular velocity $\omega$,
\begin{align}
\label{eq:dotx} \dot{x} &= v_x~,\\
\label{eq:dotvx} M \dot{v}_x & = -\frac{\partial V(x,\phi)}{\partial x} - c (x-t v_{\mathrm{s}} - x_{\mathrm{s}^0})  - \gamma M v_x ~,\\
\label{eq:dotphi} \dot\phi& = \omega ~,\\
\label{eq:dotomega} I \dot\omega &= - \frac{\partial V({x},\phi)}{\partial \phi} -\gamma I \omega~.
\end{align}

\begin{figure}
\epsfig{figure=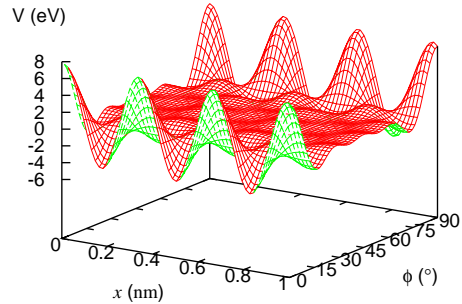,angle=270,width=7.8cm}
\vskip-\bigskipamount
\caption{
An example of the shape of $V(x,\phi)$ as obtained from the potential $V_\mathrm{A}(x,y)$ in Eq.~(\ref{eq:VA}) at constant $y$ and $z$, summed over the atoms of a nanocrystal with square contact layer of $4\times 4$ unit cells (and therefore $5\times5$ atoms) with lattice parameter 0.31585~nm on a square lattice with corrugation $V_1=1$~eV and $V_2=V_3=V_1/2$ and lattice parameter 0.32668~nm.
These parameters correspond to W(001) and NaF(001).
\label{fig:V(x,phi)}
}
\end{figure}

Some general properties of the potential energy $V(x,\phi)$ can be derived from the symmetries of the substrate lattice and contact layer.
The translation symmetry of the substrate dictates that $V(x,\phi)$ must be periodic in $x$.
A good representation of $V(x,\phi)$ is therefore given by
\begin{align}
V(x,\phi) = U(\phi) + W(\phi) \cos\left(\frac{2 \pi x}{l}\right)~,
\label{eq:potentialUW}
\end{align}
where $U(\phi)$ and $W(\phi)$ are both smooth functions that represent the average value of the potential energy and the amplitude of the modulation respectively.

Furthermore, rotational symmetries of the substrate and contact layer lead to rotational symmetries of $V(x,\phi)$ of the form
\begin{eqnarray}
V(x,\phi) = V(x, \phi_\mathrm{symmetry}+\phi)~,
\end{eqnarray}
where $\phi_\mathrm{symmetry}$ is an angle of rotation under which the contact layer is symmetric.

These symmetries of the potential $V(x,\phi)$ further imply that
\begin{align}
U(\phi) = U\left(\phi_\mathrm{symmetry}+\phi\right)~,\\
W(\phi) = W\left(\phi_\mathrm{symmetry}+\phi\right) ~.
\end{align}
For the case we consider here, square lattices, we show later in Sec.~\ref{sec:square} that also
\begin{align}
U(\phi) = U(-\phi)~,\\
W(\phi) = W(-\phi)~.
\end{align}
This also holds for some combinations of parameters of rectangular lattices.
In turn, these equations imply that $U$ and $W$ have extrema in $\phi=\phi_0=0,\phi_\mathrm{symmetry}/2$.
Since the torque, given by Eq.~(\ref{eq:dotomega}), vanishes for $\omega=0,\phi=\phi_0$ these conditions define a two-dimensional invariant manifold of the dynamics.
If the nanocrystal is on the manifold, it will remain there, and keep its orientation.
In general, there may be more invariant manifolds at other orientations, if $U$ and $W$ have additional extrema which coincide.

If the orientation of an invariant manifold is incommensurate, the corrugation is relatively small (see Fig.~\ref{fig:V(x,phi)}) and the friction force on the nanocrystal is small as well.
Such incommensurate invariant manifolds are therefore interesting for the study of nanotribology.

\begin{figure}
\epsfig{figure=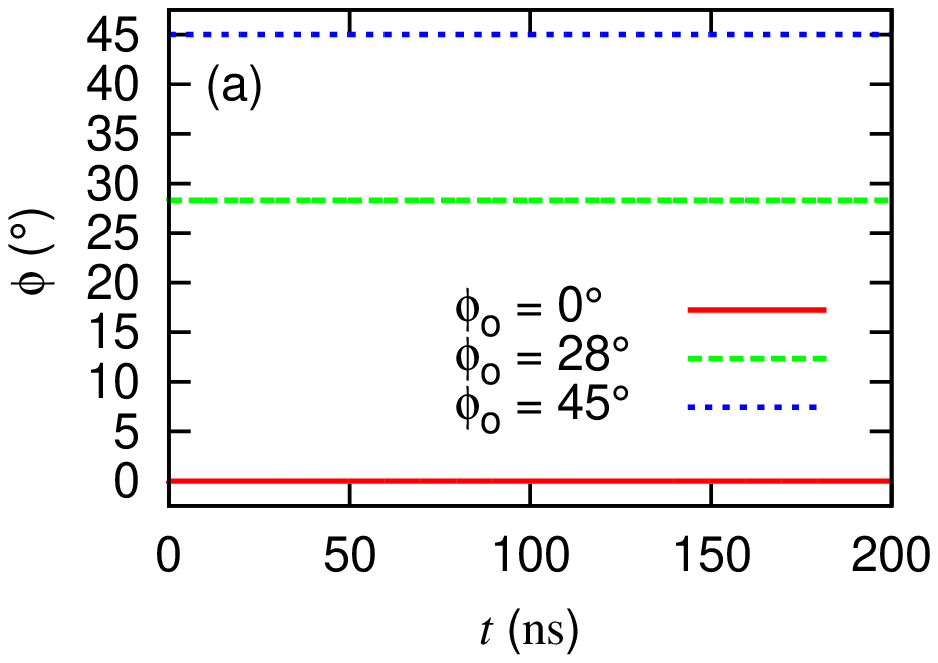,angle=270,width=4.9cm}
\epsfig{figure=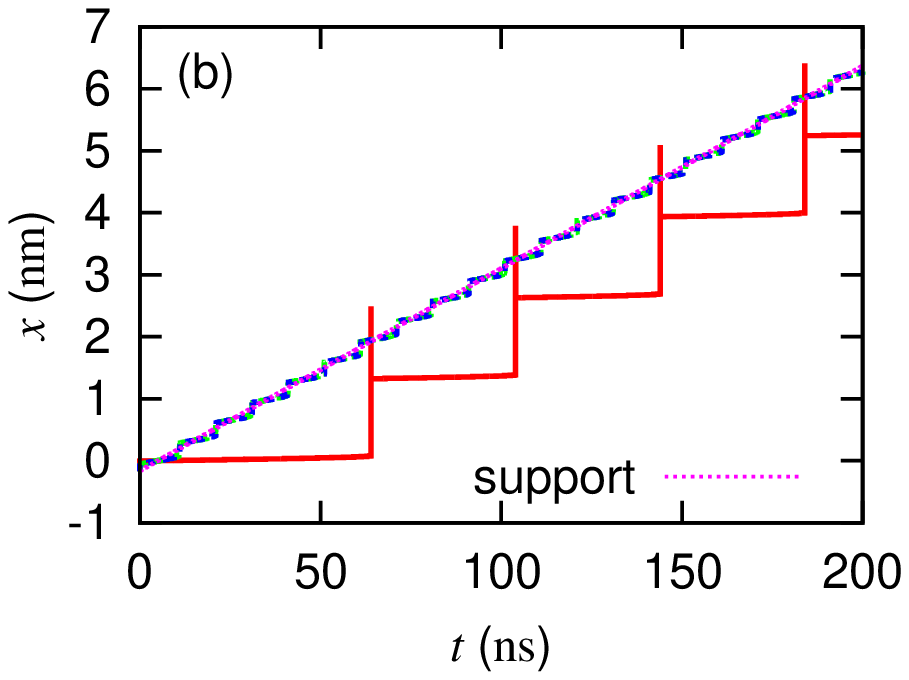,angle=270,width=4.9cm}
\epsfig{figure=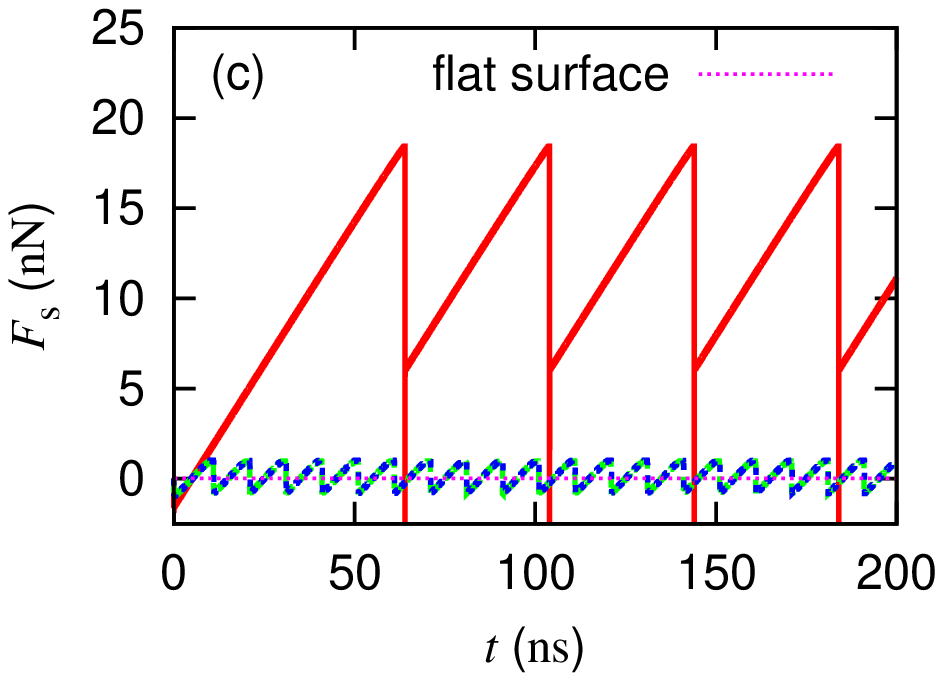,angle=270,width=4.9cm}
\caption{
Typical simulated trajectories of a W nanocrystal containing $4 \times 4 \times 4$ unit cells (contact layer $5\times5$ atoms) on NaF(001) in the absence of thermal fluctuations, at $y=a/2$, being pulled in the $x$ direction.
The orientation remains nearly constant (a) and shows only small periodic fluctuations with period $a/v_\mathrm{s}$.
Both the commensurate and incommensurate orientations display stick-slip behaviour in the position (b), but the friction (c) is significantly higher in the commensurate state.
Further details of the simulations are described in Sec.~\ref{sec:simulations}
\label{fig:sometrajectories}
}
\end{figure}

In Fig.~\ref{fig:sometrajectories}, the consequences for friction are shown for 
a typical system consisting of a W nanocrystal with (001) face on NaF(001) as described by simulations of the full two-dimensional system with the single-atom interaction potential of Eq.~(\ref{eq:VA}).
The orientation of the nanocrystal remains nearly constant and the motion in the $x$ direction exhibits stick-slip behaviour  which is relatively mild for incommensurate orientations and corresponds to a relatively low average friction force.

The existence of incommensurate invariant manifolds does not necessarily mean that low friction can be observed under experimental conditions.
The invariant manifolds may not be stable, namely, the nanocrystal may rotate away from it if it is at an orientation which deviates slightly from the invariant manifold.
We must therefore investigate the stability and robustness of the invariant manifolds.

Consider a general potential $V(x,\phi)$ which has an invariant manifold at $\phi=\phi_0$, i.e.
\begin{align}
\label{eq:phi0}
\left. \frac{\partial V(x,\phi)}{\partial \phi}\right|_{\phi=\phi_0} = 0~,
\end{align}
for all $x$.
The growth rates of perturbations in $\phi$ and $\omega$ are equal to the associated Lyapunov exponents, which are \cite{flake}
\begin{align}
\lambda_\pm = - \frac{1}{2} \gamma \pm \frac12 \sqrt{\gamma^2 - \frac{4}{I} \left\langle\left.\frac{\partial^2  V({x},\phi)}{\partial \phi^2}\right|_{\phi=\phi_0}\right\rangle_t }~,
\label{eq:lambda}
\end{align}
where $\langle \rangle_t$ denotes the time average of a quantity on a typical trajectory on the invariant manifold.
If these growth rates are smaller than zero, the invariant manifold is stable.
This is the case if the time average of the potential energy is at a local minimum, i.~e.,
\begin{align}
\label{eq:requirement}
\left\langle\left.\frac{\partial^2  V({x},\phi)}{\partial \phi^2}\right|_{\phi=\phi_0}\right\rangle_t >0~.
\end{align}

Using Eq.~(\ref{eq:potentialUW}), Eq.~(\ref{eq:requirement}) can be rewritten to read
\begin{align}
&\frac{\partial^2 U(\phi)}{\partial \phi^2}
+ \left.\frac{\partial^2W(\phi)}{\partial\phi^2}\right|_{\phi=\phi_0} \left\langle \cos\left(\frac{2 \pi x}{l}\right)\right\rangle_{t,\phi=\phi_0}
&> 0 ~.
\label{eq:requirementspecific}
\end{align}
The stability is thus determined by the functions $U$ and $W$, and how much time the particle spends near the minima of the potential, where the cosine is negative.

In stick-slip motion, the particle spends most of its time in the minima of the potential, i.e. where the cosine is smaller than zero (see Fig.~\ref{fig:sometrajectories}).
In extreme cases, $\langle \cos\rangle_t $ may be almost equal to $-1$.
If the motion is truly superlubric, then the particle spends about the same time in the minima as it does in the maxima.
If the motion is nearly superlubric, then the particle spends most of its time in the minima.
Hence, for realistic cases, $\langle \cos\rangle_t <0$.

If the offset of the potential, $U(\phi)$, has a minimum at $\phi_0$ it contributes positively towards the stability.
Similarly, if the amplitude $W(\phi)$ is at a maximum at $\phi_0$ the stability is enhanced, because the second derivative is multiplied $\langle \cos\rangle_t$, which is a negative number.
A minimum of $U$ and maximum of $W$ therefore always lead to stability, whereas a maximum of $U$ and minimum of $W$ always leads to instability.
If both are at a maximum, or both are at a minimum at $\phi_0$, then the stability is not directly obvious.

The representation of the potential used in this section not only gives qualitative understanding, but also allows for quantitative predictions about the dynamics, as is shown by comparison to full numerical simulations in Sec.~\ref{sec:simulations}.

\section{$U$ and $W$ for rectangular lattices\label{sec:UW}}

The functions $U(\phi)$ and $W(\phi)$ determine the existence and stability of the low-friction states.
We will now investigate these functions for a range of commonly occurring substrate and lattice combinations with rectangular unit cells.

\begin{figure}
\epsfig{figure=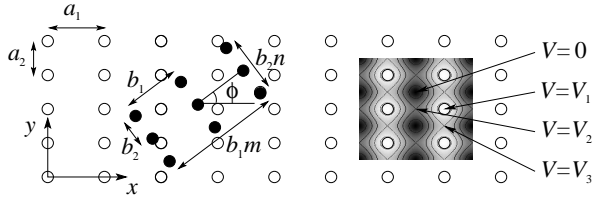,width=7.8cm}
\caption{
A top view of a general rectangular lattice (open circles) and contact layer (filled circles) with mismatch angle $\phi$, lattice parameters $a_1,a_2,b_1,b_2$, contact layer size $m,n$, and the potential energy of a contact layer atom on the substrate.
If a contact layer atom lies on top of a substrate atom, its potential energy is $V_1$.
If it lies directly between an atom and its nearest neighbour in the $x$ or $y$ direction, it has potential energy $V_2$ or $V_3$ respectively.
If it lies in the centre of a rectangle, at equal distance from four substrate atoms, without loss of generality, we may set the potential energy to 0.
The origin of the coordinate system is chosen to lie on top of a substrate atom.
}
\label{fig:lattice}
\end{figure}

Consider a two-dimensional substrate consisting of a rectangular lattice, see Fig.~\ref{fig:lattice}.
The potential energy of an atom of the contact layer at position $(X,Y)$ due to the presence of the substrate can be written in the general form
\begin{eqnarray}
\lefteqn{V_\mathrm{A}(X,Y)
 =  \frac{V_2+V_3}{2}
}&&
\nonumber \\
&&\null
+ \frac{V_1-V_2}{2} \cos \left( 2 \pi \frac{X}{a_1}\right)
+ \frac{V_1-V_3}{2} \cos \left( 2 \pi \frac{Y}{a_2}\right)
\nonumber \\
&&\null
+ \frac{V_1-V_2-V_3}{4} \left[ \cos \left( 2 \pi \frac{X}{a_1}\right) - 1\right] \left[ \cos \left( 2 \pi \frac{Y}{a_2}\right) - 1 \right]~.
\label{eq:VA}
\end{eqnarray}
Because the potential is periodic in $X$ with period $a_1$, this can be rewritten in a form similar to Eq.~(\ref{eq:potentialUW}),
\begin{eqnarray}
V_\mathrm{A}(X,Y) = U_\mathrm{A}(Y) + \mathrm{Re}\left[W_\mathrm{A}(Y) \exp\left( 2 \pi i\frac{X}{a_1}\right)\right]~,
\end{eqnarray}
with $U_\mathrm{A}(Y)$ and $W_\mathrm{A}(Y)$ the average potential energy and corrugation of a single atom travelling in the $x$ direction at constant $y=Y$.
They can be determined from the potential $V_\mathrm{A}(x,y)$ in Eq.~(\ref{eq:VA}),
\begin{eqnarray}
U_\mathrm{A}(Y) &=& \frac12 V_\mathrm{A}(0,Y) + \frac12 V_\mathrm{A}(\frac{a_1}{2},Y)\\
&= &\frac{V_1+V_2+V_3}{4}\nonumber\\
&&\null + \frac{V_1+V_2-V_3}{4} \cos \left( 2 \pi \frac{Y}{a_2}\right)~,\\
W_\mathrm{A}(Y) &=& \frac12 V_\mathrm{A}(0,Y) - \frac12 V_\mathrm{A}(\frac{a_1}{2},Y)\\
& =& \frac{ V_1 - V_2 + V_3}{4}\nonumber\\
&&\null+ \frac{V_1-V_2-V_3}{4} \cos \left( 2 \pi \frac{Y}{a_2}\right)~.
\label{eq:UA}
\end{eqnarray}

The nanocrystal is rigid, rectangular and the contact layer consists of $m \times n$ atoms.
The total potential energy of the contact layer depends on the position $(x,y)$ of the centre of mass and orientation $\phi$, and can be expressed as a sum over the potential energies of the individual contact atoms,
\begin{eqnarray}
V(x,y,\phi) = \sum_{j=1}^m \sum_{k=1}^n V_\mathrm{A}(X_{jk},Y_{jk})~,
\end{eqnarray}
where $(X_{jk},Y_{jk})$ is the position of the atom in the $j$-th column and $k$-th row of the contact layer.
For a rectangular three-dimensional nanocrystal the centre of mass is directly above the centre of the contact layer, i.~e.,
\begin{eqnarray}
&&\left(\begin{array}{l}
X_{jk} \\
Y_{jk} 
\end{array}\right)
= 
\left(\begin{array}{l}
x \\
y  
\end{array}\right)\nonumber\\
&&\null+
\left(\begin{array}{ll}
\cos\phi & -\sin\phi \\
\sin\phi & \cos\phi
\end{array}\right) \cdot
\left(\begin{array}{l}
b_1\left(j-\frac{m+1}{2}\right) \\
b_2\left(k-\frac{n+1}{2}\right) 
\end{array}\right)~.
\label{eq:rotation}
\end{eqnarray}

Let us write $V(x,y)$ in a form similar to that of Eq.~(\ref{eq:potentialUW}), as
\begin{eqnarray}
V(x,y,\phi) = U_y(\phi) + \mathrm{Re}\left[W_y(\phi) \exp\left(\frac{2 \pi i x}{a_1}\right)\right]~.
\label{eq:potentialUWY}
\end{eqnarray}
where $U_y$ and $W_y$ are the average potential energy and corrugation of a contact layer at orientation $\phi$ moving on a scan line along the $x$ direction for a specific $y$.
They can be obtained from $U_\mathrm{A}$ and $W_\mathrm{A}$ as,
\begin{eqnarray}
U_y(\phi) & = & \sum_{j=1}^m \sum_{k=1}^n U_\mathrm{A}(Y_{jk})~,\\
W_y(\phi) & = & \sum_{j=1}^m \sum_{k=1}^n W_\mathrm{A}(Y_{jk}) \exp\left( 2 \pi i \frac{X_{jk}-x}{a_1}\right)~.
\end{eqnarray}
Substituting Eq.~(\ref{eq:rotation}) and explicitly performing the sums yields
\begin{widetext}
\begin{eqnarray}
U_y(\phi) &= & m n \frac{V_1+V_2+V_3}{4}
+ \frac{V_1+V_2-V_3}{4} \cos\left(2 \pi \frac{y}{a_2}\right) \frac{\sin\left(\pi n \frac{b_2}{a_2} \cos\phi\right) \sin\left(\pi m \frac{b_1}{a_2} \sin\phi\right)}{\sin\left(\pi \frac{b_2}{a_2} \cos\phi\right) \sin\left(\pi \frac{b_1}{a_2} \sin\phi\right)}
~,
\label{eq:Urect}
\\
W_y(\phi) &=&  \frac{V_1-V_2+V_3}{4}
\frac{\sin\left(\pi m \frac{b_1}{a_1} \cos\phi\right) \sin\left(\pi n \frac{b_2}{a_1} \sin\phi\right)}{\sin\left(\pi \frac{b_1}{a_1} \cos\phi\right) \sin\left(\pi \frac{b_2}{a_1} \sin\phi\right)}
\nonumber\\ && \null
+ \frac{V_1-V_2-V_3}{4} \cos\left(2 \pi \frac{y}{a_2}\right)
\frac{\sin\left[\pi n b_2\left( \frac{\cos\phi}{a_2}-\frac{\sin\phi}{a_1}\right)\right] \sin\left[\pi m b_1\left( \frac{\cos\phi}{a_1}+\frac{\sin\phi}{a_2}\right)\right] }{\sin\left[\pi b_2\left( \frac{\cos\phi}{a_2}-\frac{\sin\phi}{a_1}\right)\right] \sin\left[\pi b_1\left( \frac{\cos\phi}{a_1}+\frac{\sin\phi}{a_2}\right)\right]}
~,
\label{eq:Wrect}
\end{eqnarray}
\end{widetext}
where we have used that $\sum_{l=0}^{d-1} \exp\{2 i x [l -(d-1)/2]\} = \sin(dx)/\sin(x)$

We consider the shape of these two functions $U$ and $W$ for the important case of square lattices in the next section.

\subsection{Square contacts on square lattices\label{sec:square}}

\begin{figure}
\epsfig{figure=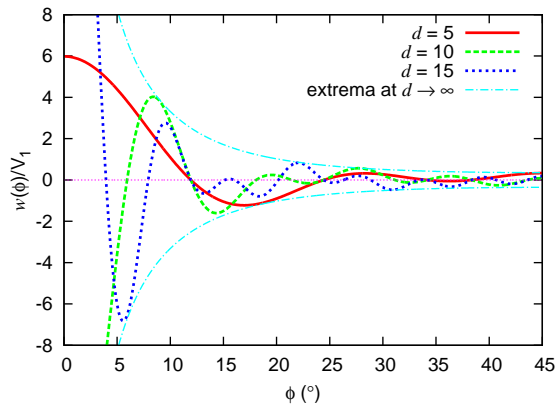,angle=270,width=7.8cm}
\caption{
The function $w(\phi)$ which determines $U_y$ and $W_y$ (see Eqs.~(\ref{eq:Uy}) and~(\ref{eq:Wyapprox}), and thus the invariant manifolds and their stability, is plotted for various values of $d$ and $b/a=0.96685$, corresponding to a W nanocrystal with its (001) face on NaF(001).
The number of extrema increases with $d$.  The maximum size of the extrema for $d \rightarrow \infty$ is also plotted.
In general, if $d$ is even, $w(\phi)$ has a minimum at 0, and if $d$ is odd a maximum.
For $d=5$, besides the maxima at $\phi=0^\circ$ and $\phi=45^\circ$, $w(\phi)$ has minima at $\phi\approx 16.95^\circ,36.25^\circ$, and a maximum at $\phi\approx 28.32^\circ$.
For $d=10$, besides the minimum at $\phi=0^\circ$ and the maximum $\phi=45^\circ$, $w(\phi)$ has minima at $\phi\approx 14.33^\circ, 22.98^\circ, 32.56^\circ, 40.86^\circ$ and maxima at $\phi\approx 8.39^\circ, 19.51^\circ, 27.69^\circ, 36.34^\circ$.
Each of the extrema corresponds to an invariant manifold.
\label{fig:UW}
}
\end{figure}

Let us consider square contact layers ($m=n\equiv d$) of atoms arranged in a square lattice ($b_1=b_2\equiv b$) on a substrate with a square lattice ($a_1=a_2 \equiv a,$ $V_2=V_3$).
For this case Eqs.~(\ref{eq:Urect}) and~(\ref{eq:Wrect}) become
\begin{eqnarray}
U_y(\phi) &= &d^2 \frac{V_1+2V_2}{4} 
\label{eq:Uy}
+ \cos\left(2\pi \frac{y}{a}\right) {w}(\phi)~,\\
W_y(\phi) & = & {w}(\phi) + \cos\left(2\pi \frac{y}{a}\right) {w}_1(\phi)~,\label{eq:Wrect2}
\end{eqnarray}
with
\begin{widetext}
\begin{eqnarray}
{w}(\phi) &=& \frac{V_1}{4} \frac{\sin\left(\pi d \frac{b}{a} \cos\phi\right) \sin\left(\pi d \frac{b}{a} \sin\phi\right)}{\sin\left(\pi \frac{b}{a} \cos\phi\right) \sin\left(\pi \frac{b}{a} \sin\phi\right)}~.\\
w_1(\phi) &=& \frac{V_1-2V_2}{4}
\frac{\sin\left[\pi d \frac{b}{a}\left(\cos\phi-\sin\phi\right)\right] \sin\left[\pi d \frac{b}{a}\left( \cos\phi+\sin\phi\right)\right] }
{ \sin\left[\pi \frac{b}{a}\left( \cos\phi-\sin\phi\right)\right] \sin\left[\pi \frac{b}{a}\left( \cos\phi+\sin\phi\right)\right]}
~.
\label{eq:Wsquareapprox}
\end{eqnarray}
\end{widetext}
These equations are symmetrical under the transformation $\phi \rightarrow -\phi$.
The second term on the right hand side of Eq.~(\ref{eq:Wrect2}), $w_1(\phi)$, is proportional to $(V_1-2V_2)/4$, which is generally small, and in many cases taken to be 0 (see, for instance, the substrate potentials used in Refs.~\cite{Hoelscher1996}).
We therefore initially neglect the second term and focus on the first term and write
\begin{eqnarray}
W_y(\phi) & = & {w}(\phi)~.\label{eq:Wyapprox}\label{eq:usquare}
\end{eqnarray}
The function $w(\phi)$ then completely determines $U$ and $W$ for all scan lines, and therefore the invariant manifolds as well as their stability.
As the first term of $U_y(\phi)$ is constant, the extrema of $U_y$ coincide approximately with the maxima of $|W_y|$, leading to invariant manifolds at the corresponding orientations,
\begin{eqnarray}
\left.\frac{\partial w(\phi)}{\partial \phi}\right|_{\phi=\phi_0} = 0~.
\end{eqnarray}
In Fig.~\ref{fig:UW} examples of $w(\phi)$ are plotted for W on NaF(001) for several values of $d$.

The extrema of $U$ and $W$ give the invariant manifolds.
One of the invariant manifolds is at $\phi=0$, while the others are at incommensurate orientations, which do not depend on the scan line $y$.
Additionally, the nodes in the amplitude ($W_y(\phi)=0$) all occur at the same value of $U_y(\phi) = d^2 (V_1+2V_2)/4$.
The number of such nodes can be determined from the number of zeros of the numerator, and in the range of $\phi \in \langle 0^\circ,45^\circ\rangle $ for $a=b$, this number is in general equal to $d-1$, except when the zeros coincide, which occurs for instance when $d^2$ can be written as the sum of the squares of two integers.

As $w(\phi)$ is independent of the position $y$ of the scan line, the orientations at which invariant manifolds occur are independent of the scan line while their stability is controlled entirely by the prefactor $\cos(2\pi y/a)$ in $U_y$.
The requirement of stability, Eq.~(\ref{eq:requirementspecific}), implies that each minimum of $U_y(\phi)$ leads to a stable invariant manifold, which in turn means that there are at least $(d-1)/2$ stable incommensurate periodic orbits if $a=b$.
If the scan line changes and the prefactor $\cos(2\pi y/a)$ becomes small, the amplitude term of $W$ in Eq.~(\ref{eq:requirement}) becomes dominant, and the invariant manifolds at the maxima of $U$ become stable as well.
If the prefactor $\cos(2\pi y/a)$ changes sign, ${U_y}$ changes sign, so that the maxima of $U$ becoming minima and the minima maxima.
The invariant manifolds which are stable for $\cos(2\pi y/a)>0$ then become unstable for $\cos(2\pi y/a)<0$, and those which were unstable become stable.
Depending on the sign and size of $\cos(2\pi y/a)$, there are therefore generally between $(d-1)/2$ and $d-1$ stable incommensurate periodic orbits.
In cases of very strong stick-slip behaviour, the centre of mass spends most of the time in the potential minimum and $\langle \cos\rangle_t $ is almost equal to $-1$. 
In this case, the invariant manifold is stable for nearly all scan lines.

When $d$ becomes large, the numerator in $w(\phi)$ oscillates rapidly.  However, the denominator is independent of $d$ and determines the size of the maxima and minima (see Fig.~\ref{fig:UW}),
\begin{eqnarray}
w_\mathrm{max}(\phi) = \frac{V_1}{4}\frac{1}{ \sin\left(\pi \frac{b}{a} \cos\phi\right) \sin\left(\pi \frac{b}{a} \sin\phi\right)}~.
\end{eqnarray}
This function therefore estimates the robustness of the incommensurate orientations, i.~e. the typical energy barrier that must be overcome to rotate from one to another nearby stable orientation,
\begin{eqnarray}
\Delta E (\phi) \approx w_\mathrm{max}(\phi)~.
\end{eqnarray}
In the case of the graphite flakes on graphite, it was found that the incommensurate orientations are not very robust \cite{flake}.
However, graphite has a very small corrugation (around 25~meV, comparable to $k_\mathrm{B} T$ at room temperature) compared to NaF(001), which has a corrugation of around 1~eV.
The typical energy barrier for W on NaF(001) is around 0.4~eV even at $45^\circ$.
Consequently, the low-friction incommensurate states of this system are robust against thermal fluctuations at room temperature.
Additionally, due to the much larger mass of the W atoms and large $M$ and $I$, the incommensurate states are also more robust against low support velocities than in graphite.

\subsection{A note on hexagonal lattices}

The results presented here can be applied to determine $U$ and $W$ for a graphite flake subject to the potential of Ref.~\cite{Dienwiebel2008}, which consists of a sum of two rectangular lattices, while a hexagonal flake can be written as the sum of several rectangular flakes.

Similar behaviour, including the existence of the transition scan line where $\cos(2\pi y/a)=0$ was observed for hexagonal flakes of different sizes on a hexagonal lattice\cite{flake}.
The diameter for hexagonal flakes is $2\sqrt{N/6}$, and the total number of invariant manifolds is $2\sqrt{N/6}+1$.
It should be noted, however, that in hexagonal lattices, due to the symmetry, the corrugation vanishes for some scan lines,
whereas this does not generally happen for square lattices.
This low $W_y$ in hexagonal lattices tends to occur at scan lines where the minimum in $U_y$ at the incommensurate orientation is not very pronounced, causing the incommensurate states to become only very weakly stable \cite{flake} and not  robust.

\section{W nanocrystals on N{a}F(001)\label{sec:simulations}}

The results of the previous section are applicable to many systems that present square lattices, like the (001) surfaces of fcc and bcc materials.
In this work, as a representative system, we have chosen a substrate which has been widely studied experimentally, NaF(001)\cite{Morita}.
Due to the bcc structure of the NaF, the surface atoms of the substrate are arranged in a square lattice with lattice parameter 0.32668~nm.
Additionally, the corrugation of NaF, around 1~eV, is typical for many substrate systems.
Unless otherwise mentioned, in this work corrugation parameters of $V_1 = 2 V_2 = 2 V_3 = 1$~eV will be used.

For the nanocrystal, we have used W, which is also common in experiments as a tip material.
W has a bcc structure and the (001) contact layer consists of a square lattice, with lattice parameter 0.31585~nm.
As the lattice parameters of the contact layer and substrate are very similar, there exist orientations at which the two lattices are nearly commensurate, as long as the contact layer is small.

This system is thus extremely suitable for studying the effects of commensurability on the friction of small nanocrystals.
As the W atoms are quite heavy, the total mass and moment of inertia are large, and the motion of the nanocrystal should behave approximately one-dimensionally and the simple model described in Secs.~\ref{sec:simple} and~\ref{sec:UW} is applicable.
The function $w(\phi)$ (Eq.~\ref{eq:Wsquareapprox}), plotted for several cases in Fig.~\ref{fig:UW} determines the stable orientations.
These orientations are mentioned in the caption for $d=5$ and $d=10$.

\subsection{simulations}

We have performed molecular dynamics simulations of rigid W nanocrystals on NaF(001) based on the full potential $V_\mathrm{A}(X,Y)$ in Eq.~(\ref{eq:VA}).
The nanocrystals were coupled harmonically (spring constant $c=10$~N/m) to a support moving at constant velocity $v_\mathrm{s}=32.668$~mm/s (or $0.1~a/\mathrm{ps}$) in the positive $x$ direction at constant $y=y_\mathrm{s}$.
Note that this is several orders of magnitude larger than the velocities typically used in AFM experiments, which are of the order of 1~$\mu$/s.
These values were chosen for computational reasons, but the dynamics of the system at lower velocities are represented well, as the time scales associated with the interaction between the substrate and nanocrystal are still much shorter than the time it takes the support to traverse one unit cell.
For the viscous friction parameter of the substrate we have chosen the typical value of $\gamma=1$/ps.
For simplicity we have restricted ourselves to cubic nanocrystals.

\begin{figure}
\epsfig{figure=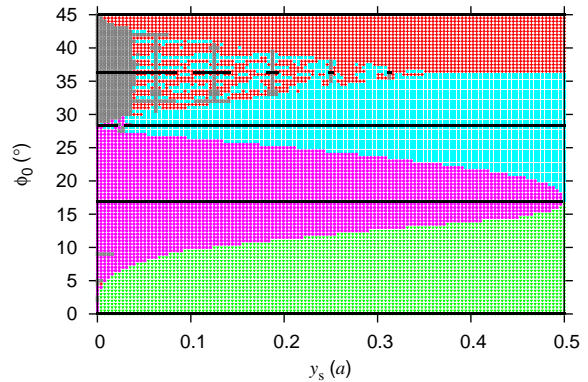,angle=270,width=7.8cm}
\caption{
\label{fig:bif5}
A bifurcation diagram for a cubic W nanocrystal with a $5\times5$ contact layer on NaF(001) at support velocity $v_\mathrm{s}=32.668$~mm/s.
Simulations were run starting from initial conditions $x=-a/4,y=y_\mathrm{s},v_x=v_\mathrm{s},v_y=0,\omega=0$ and a range of initial orientations between 0 and $45^\circ$, with $0.5^\circ$ intervals.  The final angle is plotted in black.
The set of initial angles which converge to the various periodic orbits, i.e. cross sections of the basins of attraction, are plotted in different colours for each stable orientation.
Apart from $\phi=0^\circ,45^\circ$, stable orientations occur at $\phi\approx 16.95^\circ,28.32^\circ,36.25^\circ$, which correspond exactly to the extrema of $w(\phi)$ for this system, shown in Fig.~\ref{fig:UW}.
}
\end{figure}

\begin{figure}
\epsfig{figure=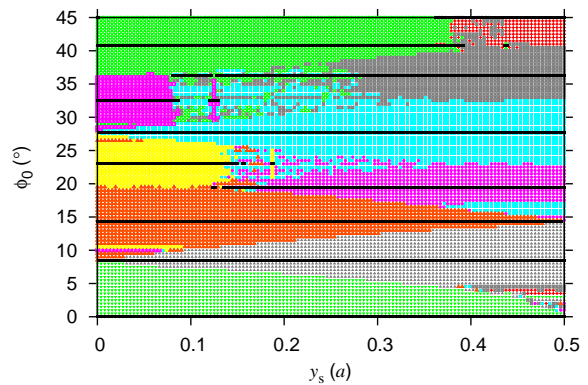,angle=270,width=7.8cm}
\caption{
A bifurcation diagram and basins of attraction of a system similar to the one of Fig.~\ref{fig:bif5} apart from the size of the W nanocrystal,
which is still cubic, but has a $10\times 10$ atom contact layer.
Apart from $\phi=0^\circ,45^\circ$, stable orientations occur at $\phi\approx 8.39^\circ, 14.33^\circ, 19.51^\circ, 22.98^\circ, 27.69^\circ, 32.56^\circ, 36.34^\circ, 40.86^\circ$.
\label{fig:bif10}
}
\end{figure}

In Fig.~\ref{fig:sometrajectories}, several examples are shown of simulated trajectories for $d=5$.
The orientations are nearly constant, but may fluctuate periodically with a period equal to the time it takes the support to move one lattice spacing.
These fluctuations are due to the motion in the $y$ direction.

In Fig.~\ref{fig:bif5} the results of 11739 simulations of the same system for different initial conditions and $y_\mathrm{s}$ are shown.
The final angles is plotted as a function of the scan line and initial angle, demonstrating the existence and stability of the periodic orbits at specific orientations.
These orientations correspond exactly to the extrema of $w(\phi)$ for this system, shown in Fig.~\ref{fig:UW}.

The sets of initial angles which converge towards these stable orientations are also indicated in Fig.~\ref{fig:bif5}.
It can be seen that the orientations at $16.95^\circ$ and $36.35^\circ$ are more stable at $y_\mathrm{s}=0$, as expected from minima of $w(\phi)$, while $0^\circ$, $28.32^\circ$, and $45^\circ$, which correspond to maxima, are, as expected, more stable at $y_\mathrm{s}=a/2$.

In Fig.~\ref{fig:bif10}, the same plot is repeated for $d=10$.
There are more stable orientations, and these correspond exactly to the extrema of $w(\phi)$ for $d=10$, shown in Fig.~\ref{fig:UW}.
Their behaviour near $y_\mathrm{s}=0$ and $y_\mathrm{s}=a/2$ is also exactly as expected from $w(\phi)$.

The orbit at $\phi=36.35^\circ$ in Fig.~\ref{fig:bif5} breaks down for some  $0<y_\mathrm{s}<a/4$.
For these scan lines the theoretical approach described in this paper predicts that this orientation is stable.
However, it is impossible to explore all possible sets of initial conditions in simulations, and thus the bifurcation diagrams are necessarily incomplete.
At these scan lines, the range of initial conditions chosen for the simulations does not intersect with the basin of attraction of the stable periodic orbit at $\phi=36.35^\circ$.

The comparison of these numerical simulations to the predictions of the analytical model is exceedingly good, including the important feature that several incommensurate orbits are stable for all scan lines.

\begin{figure}
\epsfig{figure=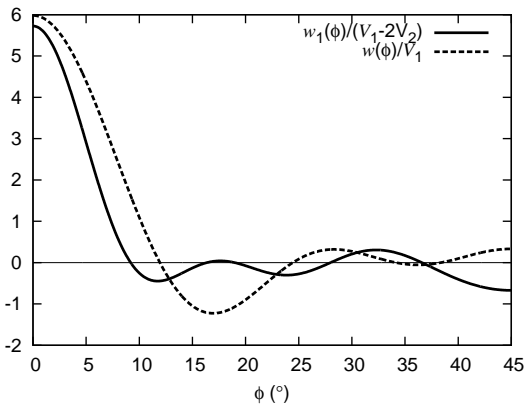,angle=270,width=7.8cm}
\caption{
The functions $w(\phi)$ and $w_1(\phi)$ are plotted for $b/a=0.96685$, corresponding to a system of W(001) on NaF(001) for $d=5$ and $V_2=V_1/3$.
$w(\phi)$ is the same as in Fig.~\ref{fig:UW}, but $W_y(\phi)$ is no longer equal to it.
Besides the maximum at $\phi=0^\circ$ and minimum at $\phi=45^\circ$, $W_y(\phi)$ has minima at $\phi\approx 11.73^\circ,23.83^\circ$, and maxima at $\phi\approx 17.63^\circ, 32.39^\circ$.
\label{fig:UWasym}
}
\end{figure}

\begin{figure}
\epsfig{figure=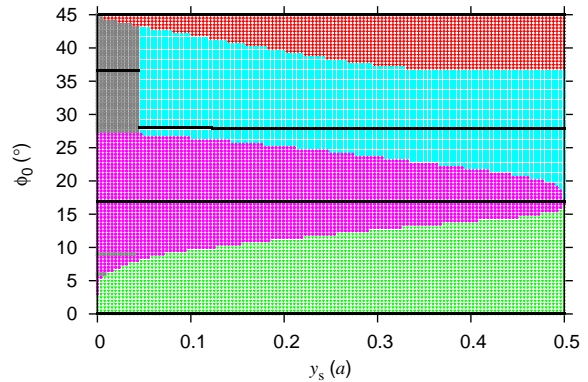,angle=270,width=7.8cm}
\caption{
\label{fig:bif5asym}
The plot of Fig.~\ref{fig:bif5} repeated with $V_2=V_1/3$.
Besides at $\phi=0^\circ,45^\circ$, stable orientations occur at $\phi\approx 16.89^\circ,27.9^\circ,36.59^\circ$.
}
\end{figure}

\subsection{The case $V_2 = V_3 \neq V_1/2$}

Though in most cases of square lattices $V_2 = V_1/2$ (see, for instance Ref.~\cite{Hoelscher1996}), this may not always be true.
We therefore briefly consider the implications of $V_2=V_3\neq V_1/2$, where, in Eq.~(\ref{eq:Wrect2}), the second term can no longer be neglected.
In Fig.~\ref{fig:UWasym} $w(\phi)$ and $w_1(\phi)$, which determine $U_y(\phi)$ and $W_y(\phi)$ through Eqs.~(\ref{eq:Uy}) and~(\ref{eq:Wrect2}) are plotted.
These two functions together determine the invariant manifolds and their stability.
The extrema of $U$ and $W$ no longer coincide exactly, and so the location of the invariant manifolds is much less trivial than before.

In Fig.~\ref{fig:bif5asym}, the plot of Fig.~\ref{fig:bif5} is repeated for $V_2=V_1/3$.
As both $w(\phi)$ and $W_y(\phi)$ still have an extremum near $\phi\approx17^\circ$, the orbit at that orientation persists, but is changed slightly.
The orbits near $28^\circ$ and $36^\circ$ also still exist, in distorted form, because the second term in $W_y$, determined by $w_1(\phi)\propto (V_1-V_2)/V_2$, is relatively small compared to $U_y(\phi)$, so that $U_y(\phi)$ is the dominant contribution to the stability, regardless of the scan line.
This is in general the case, unless $V_2$ deviates very strongly from $V_1/2$.

\section{Conclusions\label{sec:discussion}}

We have analytically investigated the driven nonlinear dynamics of general rectangular nanocrystals on rectangular lattices and their relation to sliding friction.
We have formulated an approximate analytical model, which gives the conditions for the existence and stability of incommensurate sliding states with low friction.
We show that the number of incommensurate orbits grows linearly with the diameter of the nanocrystal and that for realistic systems, several states can be robust against thermal fluctuations, change of scan line, and driving velocity.
The geometry and high typical corrugation of substrates with square lattices increase the robustness compared to typical hexagonal lattices, such as graphite.
Moreover, unlike hexagonal lattices, square lattices do not have a scan line where the stability becomes extremely weak.
This has implications for experiments where many different scan lines are explored, and thus a nanocrystal at a stable orientation for one scan line would be forced to rotate when the apparatus switches to another scan line, as was noted for the specific case of graphite flakes on graphite in Ref.~\cite{flake}.

Together with the increased moment of inertia for a nanocrystal, our results suggest that it should be easier to experimentally observe incommensurate sliding for three-dimensional nanocrystals with bcc or fcc structure than for hexagonal graphite flakes.

\begin{acknowledgements}
ASW's work is financially supported by a Veni grant of Netherlands Organisation for Scientific Research (NWO).
\end{acknowledgements}

\end{document}